# Solar Power Systems Web Monitoring

Bimal Aklesh Kumar
Department of Computer Science and Information Systems
Fiji National University

*Abstract* — All over the world the peak demand load is increasing and the load factor is decreasing year-by-year. The fossil fuel is considered insufficient thus solar energy systems are becoming more and more useful, not only in terms of installation but monitoring of these systems is very crucial. Monitoring becomes very important when there are a large number of solar panels. Monitoring would allow early detection if the output falls below required level or one of the solar panel out of 1000 goes down. In this study the target is to monitor and control a developed solar panel by using available internet foundation. This web-enabled software will provide more flexibility over the system such as transmitting data from panel to the host computer and disseminating information to relevant stake holders barring any geographical barrier. The software would be built around web server with dynamic HTML and JAVA, this paper presents the preliminary design of the proposed system.

## I. INTRODUCTION

Solar energy, radiant light and heat from the sun, has been harnessed by humans since ancient times using a range of ever-evolving technologies. Solar radiation, along with secondary solar-powered resources such as wind and wave power, hydroelectricity and biomass, account for most of the available renewable energy on earth. Only a minuscule fraction of the available solar energy is used [1].

Solar powered electrical generation relies on heat engines and photovoltaic. Solar energy's uses are limited only by human ingenuity. A partial list of solar applications includes space heating and cooling through solar architecture, potable water via distillation and disinfection, day lighting, solar hot water, solar cooking, and high temperature process heat for industrial purposes. To harvest the solar energy, the most common way is to use solar panels.

Solar technologies are broadly characterized as either passive solar or active solar depending on the way they capture, convert and distribute solar energy. Active solar techniques include the use of photovoltaic panels and solar thermal collectors to harness the energy. Passive solar techniques include orienting a building to the Sun, selecting materials with favorable thermal mass or light dispersing properties, and designing spaces that naturally circulate air [1].

Not only installation of these technologies but also monitoring and controlling of these systems are very important. A study by solar experts concluded that about half of all solar power systems aren't working as they should, and this leads to around 20% of a year's solar electricity to be lost [2].

In this study we aim to monitor and control solar panels using web based system. The wide spread use of web has removed distance and communication barriers. The system would be developed using HTML, Java, and JSP it would provide various techniques to remotely manage solar panels set up at a single home or likewise a large solar farm.

## II. RESEARCH BACKGROUND

Micro-inverters (and other parallel technology) were given lots of attention because they can increase the efficiency of a system by up to as much as 10%-20%. Similarly, solar electricity systems that are hooked up to monitoring systems have a 10% energy production increase over systems that are not hooked up to monitoring systems, according to Will Shortt, CEO of Deck Monitoring [4].

PV solar panels last at least 25 years, whereas inverters only come with an 8-10 year warranty. That means that sometime in the 8-10 year range the inverter will die and the system will stop producing energy. With a monitoring system in we can know immediately that the system has been compromised. Otherwise it could be weeks or months before that we realize that the solar electricity system is not longer producing energy [3].

With a monitoring system in place an installer could offer a "performance assurance", and that may be just the differentiator needed to close the deal similarly, it may motivate installers to offer warranties, with monitoring systems in place they clean the panels when they see the productivity drop rather than the other extreme of having a system fail for some reason.

The most compelling feature of monitoring systems is the ability to measure performance against what was promised and what is expected of the system. Not only is it fun to see, but it also serves as a great indicator if something goes wrong with the system.

Lastly, when owners are able to view their solar energy production or energy usage in a clear, easy to view fashion, they inevitably will adjust their behavior and start using less energy, which is a great positive side-effect.

## III. WEB BASED MONITORING

In order to make a web based monitoring system, we have to use the following technology; HTML, Java and web server. In order to observe and control the system while working one computer has been used and already present and LAN line is connected to the internet.

**HTML** - which stands for Hypertext Markup Language, is the predominant markup language for web pages. HTML elements are the basic building-blocks of webpages. HTML is written in the form of HTML elements consisting of tags, enclosed in angle brackets (like html>), within the web page content. HTML tags normally come in pairs like <h1> and </h1>. The first tag in a pair is the start tag, the second tag is the end tag (they are also called opening tags and closing tags). In between these tags web designers can add text, tables, images, etc.

**Java -** is a programming language originally developed by James Gosling at Sun Microsystems (which is now a subsidiary of Oracle Corporation) and released in 1995 as a core component of Sun Microsystems' Java platform. The language derives much of its syntax from C and C++ but has a simpler object model and fewer low-level facilities. Java applications are typically compiled to byte code (class file) that can run on any Java Virtual Machine (JVM) regardless of computer architecture.

**Java Server Pages (JSP)** - JSP technology enables web developers and designers to rapidly develop and easily maintain, information-rich, dynamic web pages that leverage existing business systems.

**Web Server** - can refer to either the hardware (the computer) or the software (the computer application) that helps to deliver content that can be accessed through the Internet.[1] The most common use of web servers is to host web sites but there are other uses like data storage or for running enterprise applications. The primary function of a web server is to deliver web pages on the request to clients. This means delivery of HTML documents and any additional content that may be included by a document, such as images, style sheets and scripts.

**Data Logger** is an electronic device that records data over time or in relation to location either with a built in instrument or sensor or via external instruments and sensors. Increasingly, but not entirely, they are based on a digital processor (or computer). They generally are small, battery powered, portable, and equipped with a microprocessor, internal memory for data storage, and sensors. Some data loggers interface with a personal computer and utilize software to activate the data logger and view and analyze the collected data.

**Database** is an organized collection of data for one or more purposes, usually in digital form. The data are typically organized to model relevant aspects of reality (for example, the availability of rooms in hotels), in a way that supports processes requiring this information (for example, finding a hotel with vacancies). The term "database" refers both to the way its users view it, and to the logical and physical materialization of its data, content, in files, computer memory, and computer data storage.

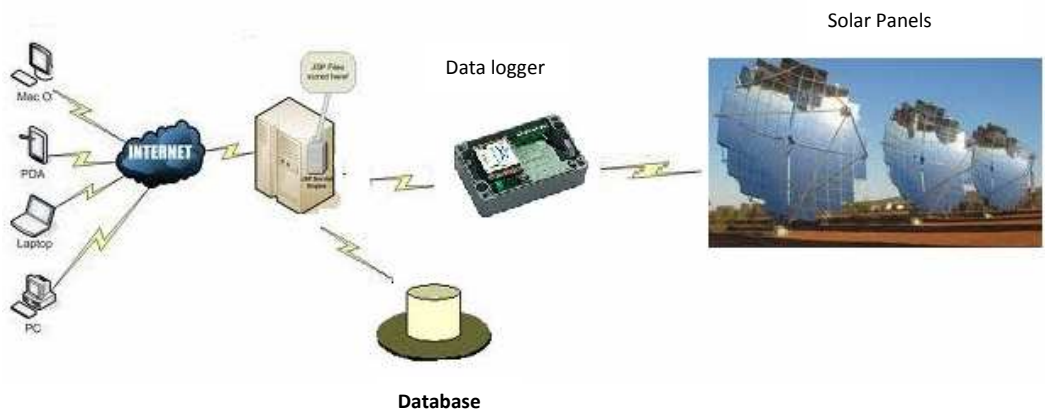

Fig. 1 General Structure of Solar Power Web Monitoring Systems

Figure 1. above gives the basic structure of web based monitoring systems. The data logger reads data from solar panels and stores information in storage device. The web server reads the data from the storage device the clients can access the information on solar panels using internet.

The benefits of web based monitoring systems include;

1. Maintenance and reparation will become much easier by means of utilization of internet sub structure.
2. Java program is accepted on many platforms thus there is no need to select a platform for installation.
3. The long range connections can be operated with the present structure without any extra hardware and software needed.
4. Its maintenance and reparation will be easy due to its limited number of connections.

To view the energy yield of our systems, a graphic program will installed web browser will be able to query data from the database and display the desired energy graphs a sample is given below.

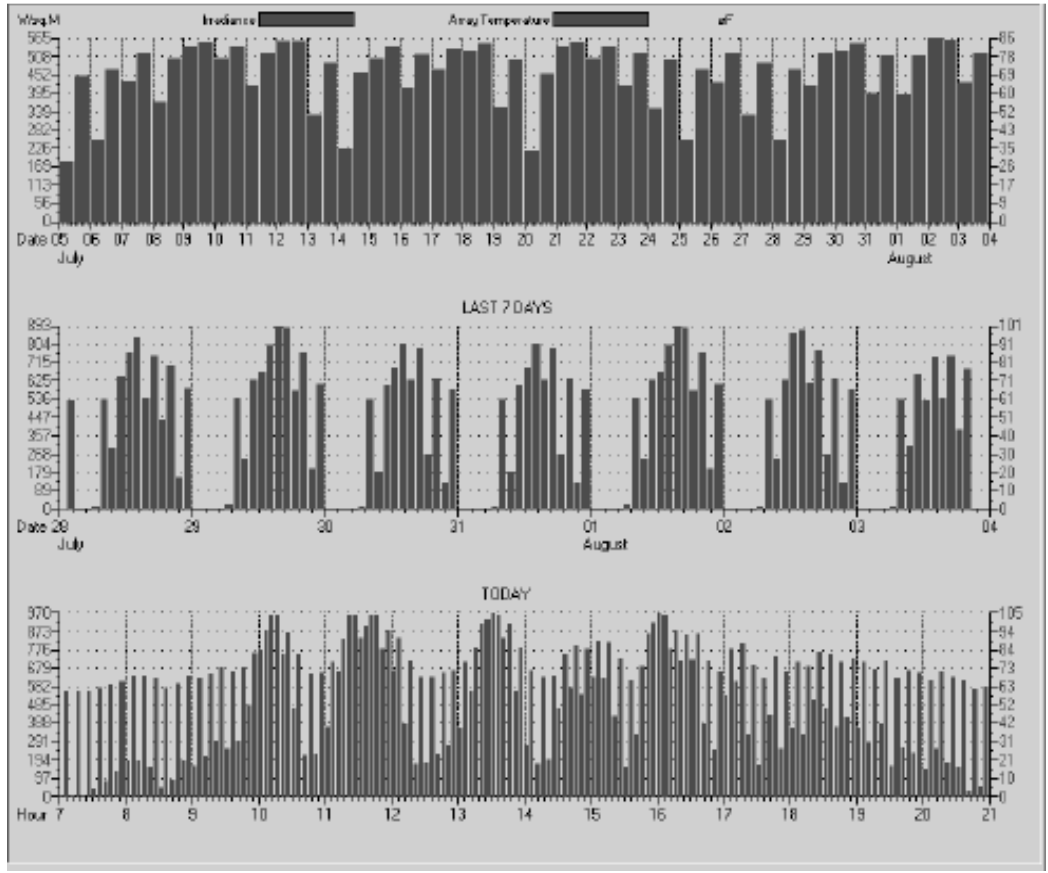

Fig. 2 Energy yield graph

CONCLUSIONS AND FUTURE WORK

This paper outlined plans for the implementation of web based monitoring of solar panels and it benefits. Future work would be carried out for the implementation of this system.

REFERENCES

[1] http://en.wikipedia.org/wiki/Solar_energy
[2] http://solarnowsystems.com/?page_id=882
[3] S. Abdallah (2008), Sun tracking system for productivity enhancement of Solar. Still Desalination pp. 669-676.
[4] http://www.residentialsolar101.org/solar-energy-monitoring-systems.

**Bimal Aklesh Kumar** is a lecturer at Fiji National University for past seven years in the department of Computing Science and Information Systems. His research interest includes software engineering, distributed systems and internet computing. Mr. Kumar has presented two papers at international and national conferences and published three journal articles in the field of software engineering.